\documentclass[twocolumn]{revtex4-1}
\usepackage{epsfig}
\usepackage{graphics}
\usepackage{amssymb}
\usepackage{textcomp}
\usepackage{amsmath}
\usepackage[dvipsnames]{xcolor}
\usepackage{subfigure}

\def\eq#1{(\ref{#1})}

\begin{document}

\title{Spin networks and the big bang singularity avoidance in the AdS/CFT correspondence}

\author{Carlos Silva} 
\email{carlosalex.phys@gmail.com}
\affiliation{Instituto Federal de Educa\c{c}\~{a}o Ci\^{e}ncia e Tecnologia do Cear\'{a} (IFCE),\\ Campus Tiangu\'{a} - Av. Tabeli\~{a}o Luiz Nogueira de Lima, s/n - Santo Ant\^{o}nio, Tiangu\'{a} - CE, Brasil}

\date{\today}

\begin{abstract}

We propose a new version of the $AdS/CFT$ correspondence, where polymer structures similar to Loop Quantum Cosmology spin networks can be induced on an isotropic and homogeneous flat  Randall-Sundrum II brane, corresponding to the holographic duals of closed string states living in the bulk.
Such polymer structures drive a discrete evolution of the braneworld spacetime that leads to the regularization of the Big Bang singularity.
The present results show that the $AdS/CFT$ conjecture can be a possible bridge between String theory and Loop Quantum Gravity.

\end{abstract}


\maketitle

Amidst the difficulties to obtain a quantum description of gravity,  the holographic principle emerges as a possible guide
\cite{'tHooft:1993gx, Susskind:1994vu}. Such principle, which has been supported by observational evidence  \cite{Afshordi:2016dvb}, appears in its most successful form, as the so-called $AdS/CFT$ conjecture, which  has established that a gravitational theory in the $AdS_{5}$ bulk is dual to a CFT with a UV cut-off on the bulk boundary \cite{Maldacena:1997re}. 
%Particularly, if such boundary is a Randall-Sundrum II brane, the CFTs must be coupled to gravity \cite{Randall:1999vf, Duff:2000mt}. 

%The issue of the Big Bang singularity has been addressed in  scenarios proposed by String theory \cite{Gasperini:2002bn}, and by ekpyrotic and cyclic models \cite{Khoury:2001wf, Khoury:2001bz} where, although we have the possibility of a pre-big-bang phase to our universe, we do not have a deterministic way to join the pre-big-bang phase to the present post-big-bang era. It occurs because these treatments are perturbative and depend on the existence of a background that is a smooth continuum. 
 
As a definition of nonperturbative string theory, the $AdS/CFT$ correspondence has offered the interesting possibility to solve gravitational singularities, by transfer the discussion from the bulk to the dual field theory on its boundary. However, in the important case of the Big Bang singularity resolution, no clear picture has emerged by now in this context, despite much effort has gone. See e.g. \cite{Engelhardt:2014mea, Engelhardt:2015gta, Bodendorfer:2016gba, Bodendorfer:2018dgg} for more recent discussions on this issue. Particularly,  in \cite{Bodendorfer:2016gba, Bodendorfer:2018dgg} Loop Quantum Gravity (LQG) ideas have been used to this purpose.

%However, despite some efforts \cite{Craps:2007ch}, the application of the $AdS/CFT$ conjecture to the issue of the Big Bang still run in the problem of the break down of the dynamical equations at the initial singularity.  In fact,  it occurs because these treatments are perturbative and depend on the existence of a background that is a smooth continuum. 

%Such difficulties are shared by others scenarios proposed by String theory \cite{Gasperini:2002bn}, and by ekpyrotic and cyclic models \cite{Khoury:2001wf, Khoury:2001bz} where, although we have the possibility of a pre-big-bang phase to our universe, we do not have a deterministic way to join the pre-big-bang phase to the present post-big-bang era.

The situation becomes even more cloudy if we consider that the bulk boundary consists of a Randall-Sundrum II brane, where the $CFT$ is coupled to gravity \cite{Randall:1999vf, Duff:2000mt}. In fact, in such a scenario, we have no idea what the boundary theory would be.  Notwithstanding, an important hint has been pointed out that, whatever such theory is, it should be hot and have a non-zero energy density and pressure \cite{Savonije:2001nd}. It is because bulk black holes could emit Hawking radiation that would heat the brane, which by its turn, would be forced by the Israel equations to turn into an Friedmann-Robertson-Lama\^{i}-Walker (FRLW) universe driven by an energy density that dilutes as a finite temperature CFT.

More recently, such a thermodynamical aspect of the boundary theory has been reinterpreted by an interesting perspective that arises when we locate the black holes on the brane itself. In this context, the thermal radiation emitted by such holographic black holes provides the thermal nature of the boundary theory, assuming an interesting place in the $AdS/CFT$ dictionary. In fact, this radiation corresponds to the holographic dual of gravitational bremsstrahlung in the $AdS_5$ bulk \cite{Tanaka:2002rb, Emparan:2002px, Anderson:2004md}. Such results bring us a reformulation of the $AdS/CFT$ correspondence in the form of the so-called Black Hole Holographic Conjecture (BHHC).

In the present work, we shall demonstrate that the BHHC can be used to introduce a regularization scheme whereby the Big Bang singularity can be solved, in an $AdS/CFT$ scenario where the boundary theory lives on an  isotropic and homogeneous flat brane universe. In such a regularization scheme, the gravitational degrees of freedom belonging to the theory on the boundary can be associated with polymer structures similar to Loop Quantum Cosmology (LQC) spin networks. Such structures will
appear as the holographic duals of string states in the bulk and turn the Big Bang singularity into a bounce.
Such results show that the $AdS/CFT$ conjecture can be a possible bridge between string theory and LQG, the two main approaches to quantum gravity up to now.

%In the present work, we show that more light can be shed on the nature of the boundary theory, through a reformulation of the $AdS/CFT$ correspondence, where the gravitational degrees of freedom on the brane can be associated with polymer structures similar to Loop Quantum Cosmology (LQC) spin networks. Such structures will
%appear as the holographic duals of string states in the bulk, and will drive a cosmological evolution of the brane in a way that the Big Bang singularity can be solved in this context.  

%The present results can establish a possible connection between String theory and Loop Quantum Gravity (LQG), which have been considered as the main candidates to a theory of quantum gravity up to now.

\vspace{2mm}

\noindent \emph{\textbf{- Spacetime degrees of freedom on a brane: a BHHC perspective:}}

\vspace{2mm}

%In order to solve the Big bang singularity in a Randall-Sundrum II scenario, we shall  demonstrate that the BHHC implies in a redefinition of the number of degrees of freedom related with the gravitational theory induced on the brane, in a way that we can regularize the Big Bang/ Big Crunch singularit in the $AdS/CFT$ context.

In field theory, a regularization scheme consists of the determination of degrees of freedom, in such a way, those that lead to divergences are cut out from the theory \cite{Weinberg:1995mt}. 
In this sense, by considering the Bekenstein bound \cite{Bekenstein:1980jp}, the discussion about the degrees of freedom present in the theory induced on a brane must be connected with
what $4D$ solution, describing a black hole, we have localized on it.
Such a solution must be given by a suitable slicing of a $5D$ accelerating
black hole metric, known as the C-metric in $4D$ \cite{Kinnersley:1970zw}.

Several approaches have been proposed in the literature, where induced $4D$ black hole metrics of the form

\begin{equation}
ds^{2} = - F(r)^{2}dt^{2} + G(r)^{2}dr^{2} + H(r)^{2}d\Omega^{2} \label{g-gorm} 
\end{equation}

\noindent has been used \cite{Chamblin:1999by, Dadhich:2000am, Casadio:2001jg, Visser:2002vg, Bronnikov:2003gx}. Unfortunately, none of these proposals has been satisfactory so far.

Among the points to be clarified, we note that all the approaches to construct black hole solutions on a brane
have the common assumption that the radial coordinate must be fixed by the "`area gauge"' $H(r) = r$. Consequently, the area of the $2$-spheres surrounding the black hole behaves as
$A(r) = 4\pi r^{2}$, increasing monotonically
between the horizon and the spacelike infinity.

On the other hand, a different perspective has been driven by a more recent version of the AdS/CFT conjecture, proposed by Tanaka \cite{Tanaka:2002rb}, and independently by Emparan, Fabbri and Kaloper \cite{Emparan:2002px}, which states that for regimes above the $AdS$ length scale:

\vspace{5mm}

\noindent \emph{ A classical $5$-dimensional gravitational theory in the bulk corresponds to a quantum corrected (semiclassical) 4D braneworld spacetime.}
\vspace{5mm}

\noindent This new perspective has been named by Gregory et al as the "Black Hole Holographic Conjecture" (BHHC) \cite{Gregory:2004vt}.

In the context of such idea, black hole solutions localized on the brane,
given as solutions of the classical bulk equations in $AdS_{D+1}$ with the
brane boundary conditions, can not be static but must consist in evaporating solutions, corresponding to quantum-corrected black holes in
$D$ dimensions, rather than classical ones. In this way, in the $AdS/CFT$ dictionary, extra-dimensional effects in the bulk will be translated into semiclassical corrections
to the boundary theory. Some possibilities for observational confirmation of the BHHC has been pointed in \cite{Emparan:2002jp, Dey:2020lhq}.

The use of the BHHC has led to the idea that
the area function $A(r)$ of the $2$-
spheres, in the geometry induced on the brane, must be considered to be not monotonic \cite{Gregory:2004vt}.
The main reason for such an assumption comes from
the presumed higher dimensional $C$-metric. Such a metric would consist of an accelerating
black hole being "`pulled" by a cosmic string. Since the appropriate higher-dimensional metric
for a Poincar\'{e} invariant string has a turning point in the area function, and the
"horizon" is singular, it is expected that a braneworld black hole solution must share such features.
A turning point in the area function of a holographic black hole also appears in \cite{Casadio:2004nz}, where it has been demonstrated that, due to back-reaction effects, an anti-evaporating phase must take place at the end of the black hole evaporation process. 

It is possible to introduce a scenario, as prescribed  by the BHHC,  by considering a fluctuation of a black hole metric written in the area
gauge $d{s}_{ag}^{2}$ in the following way

\begin{equation}
ds^{2} = \frac{H^2}{r^2} d{s}_{ag}^{2} \; ,  \label{resc-eq}
\end{equation}

\noindent where we must consider an appropriate choice of the function $H(r)$. In this case, if  $H/r$ is a
twice differentiable function of the spacetime coordinates, and $0 < H/r <  \infty$, we can interpret the equation above as a conformal transformation.

To obtain the desired scenario, a possible choice is:

\begin{equation}
H^2(r) = \Big(r^{2}+
\frac{l^{4}}{r^{2}}\Big) \label{conformal-factor} \; , 
\end{equation}

\noindent in a way that the metric $ds^{2}$ will possess a not monotonic area function, with a turning point at $r = l$, and singular at $r= 0$. In this scenario, the length $l$ will correspond to a high energy length scale and will enclose the information about higher-dimensional effects, translated as semiclassical corrections on the brane. A similar conformal factor has been used in \cite{Bambi:2016wdn, Bambi:2016yne} to solve the Schwarzschild black hole singularity.
%We note yet that such scenario corresponds to that where the black hole must enter
%in an anti-evaporation phase, when the area function reaches its turning point, a phenomena that has been foreseen for holographic black holes \cite{Casadio:2004nz}.

Interestingly, due to the conformal relationship between the metrics $ds^{2}$ and $d{s}_{ag}^{2}$, the way how the black hole temperature and entropy are related with the black hole mass is the same
in both spacetimes \cite{Bambi:2016yne}. However, since the event horizon surface area $A$ changes by conformal transformations, the black hole entropy-area relation
must be modified.
In fact, by considering the Eq. \eq{conformal-factor}, a new radial coordinate in the $ds^{2}$ spacetime will be defined as

\begin{equation}
R = \Big(r^{2}+
\frac{l^{4}}{r^{2}}\Big)^{1/2} \label{phys-rad}\; .
\end{equation}

\noindent Consequently, the rescaled event horizon area will be given by

\begin{equation}
A = 4\pi\Big(r_{+}^{2}+
\frac{l^{4}}{r_{+}^{2}}\Big)\;\; , \label{r-area}
\end{equation}

\noindent where $r_{+}$ is the black hole event horizon radius in the $d{s}_{ag}^{2}$ spacetime.

We can fix the bare metric $d{s}_{ag}^{2}$ in a way that the spacetime described by it obeys the usual Bekenstein-Hawking entropy-area relation, in four dimensions \cite{Bekenstein:1973ur} (see, for example, the general class of braneworld black holes introduced in
\cite{Abdalla:2006qj}). In this way, we obtain for the $ds^{2}$ spacetime:

\begin{eqnarray}
S  &=& \frac{A + \sqrt{A^{2} - 2A_{l}^{2}}}{8} \nonumber \\
    &=& \;\;\; \frac{\sqrt{A^{2} - A_{l}^{2}}}{4} +  \mathcal{O}(\geq A_{l}^{6}) \;\;\; , \label{entropy-area}
\end{eqnarray}

\noindent where $A_{l} = 4\sqrt{2\pi} l^{2}$.

In the present work, we shall consider that the presence of a turn-point in the area function of a black hole, in the context of the BHHC, allow us to introduce a cut-off on the degrees of freedom that can be encoded on the bulk boundary, in a way that their counting will be given by the expression \eq{entropy-area}. Such a cut-off will be implemented through the parameter $A_{l}$.

We must cite that the expression \eq{entropy-area} has also appeared in the context of loop quantum black holes \cite{Modesto:2009ve}, and has been used to derive the LQC dynamical equations from thermodynamical arguments \cite{Silva:2015qna}.
Moreover, the present considerations must be valid for regimes above the $AdS$ length scale. Below such scale, all dimensions must be seen on the same footing and the full influence of the extra dimensions must be taken into account. Other choices for the function $H(r)$ can be done in order to implement regularization schemes different from that we have adopted here.

\vspace{2mm}

\noindent \emph{\textbf{ - Cosmology and the Big Bang singularity avoidance on a  brane:}}
\vspace{2mm}

%In the $AdS/CFT$ scenario, if we consider an asymptotically AdS space-time contracting towards a
%curvature singularity, the bulk theory cannot be used to evolve further
%in time. Consequently, a natural route of investigation is to analyse if it may be possible to track the time evolution in the dual field theory in a controlled fashion, i.e, if  the dual  theory admits a continuation in time beyond to that the bulk singularity occurs \cite{Engelhardt:2014mea, Engelhardt:2015gta, Bodendorfer:2016gba, Bodendorfer:2018dgg}.

%%%%%%%%%%%%%%%%%%%%%%%%%%%%%%%%%%%%%%%%%%%%%%
%%%%%%%%%%%%%%%%%%%%%%%%%%%%%%%%%%

To address the cosmological evolution of the brane spacetime in the light of the BHHC, let us consider the flux of Hawking radiation through the universe horizon.
For this flux, we can associate the following temperature
\cite{Ge:2007yu, Gong:2007md, Cai:2008gw}:

\begin{equation}
T = (2\pi\tilde{r}_{H})^{-1}\;\;. \label{kodama-temperature}
\end{equation} 

\noindent In the expression above, $\tilde{r}_{H}$ is the physical radius associated with the universe horizon,  given for the flat universe case by

\begin{equation}
\tilde{r}_{H} = \frac{1}{H}\; , \label{ap.hor-rad}\;
\end{equation}

\noindent where we have considered an FLRW metric induced on the brane:

\begin{eqnarray}
ds_{FRLW}^{2} &=& h_{ab}dx^{a}dx^{b} + \tilde{r}^{2}d\Omega^{2}_{2}\label{frw-metric} \; ,
\end{eqnarray}

\noindent with $h_{ab} = diag (-1, a^{2})$, $\tilde{r} = a(t)r$, and $H = \dot{a}/{a}$.

As usual, we shall consider that the energy-momentum tensor of the matter-energy contend
of the boundary theory can be written as the ones for a perfect
fluid, which gives us an amount of energy that goes through
the horizon during a time $dt$ as \cite{Cai:2005ra}:

\begin{eqnarray}
dQ = A\psi =  A(\rho+p)\tilde{r}H\Big(1+ \frac{\rho}{\sigma}\Big)dt \; ,
\end{eqnarray}

\noindent where $A = 4\pi\tilde{r}^{2}_{H}$, and $\sigma$ is the brane tension.

Now, we observe that from a global point of view, the universe horizon enters in the bulk, in a way that higher-dimensional effects must be taken into account to the calculation of its entropy. However, according to BHHC, for regimes above the $AdS$ length scale,
such effects can be interpreted as semiclassical corrections on the brane. In this case,  we shall borrow the expression  \eq{entropy-area} from the last section, and use it to express the entropy associated with the universe horizon.

In this way, by taking into account
the first law of thermodynamics,
$dQ = TdS$, and the temperature \eq{kodama-temperature},
we obtain, by discarding higher-order corrections in $A_{l}$:

\begin{equation}
\dot{H}  =   4\pi  \frac{\sqrt{A^{2}-A_{l}^{2}}}{A}(\rho + p)\Big(1+ \frac{\rho}{\sigma}\Big) 
\;,
\label{friedmann1}
\end{equation}

\noindent which, by the use of the continuity equation, give us:

\begin{equation}
\frac{8\pi}{3}\frac{d\rho}{dt}\Big(1+ \frac{\rho}{\sigma}\Big)  = 
\frac{A}{\sqrt{A^{2} - A_{l}^{2}}} \frac{d(H^{2})}{dt} \; . \label{friedmann1.1}
\end{equation}

\noindent Finally, by integration, we obtain the Friedmann equation:

\begin{equation}
H^2  = \frac{4\pi}{A_{l}}\cos(\Theta) \;, \label{friedmann2}
\end{equation}

\noindent where $\Theta = \pm\Big[\frac{2A_{l}}{3}\rho\Big(1+ \frac{1}{2}\frac{\rho}{\sigma}\Big) - \alpha\Big]$, and 
$\alpha$ is a phase 
constant. 

Note that, in the Eq. \eq{friedmann2}, the effective density term,
in the form of a harmonic function of the classical density, bring us a scenario where a bounce takes the place of the Big Bang initial
singularity, when the universe density assumes a critical value, differently from usual braneworld cosmology \cite{Maartens:2003tw}.

To explore the physical significance of the result above, let us expand the right-hand side of the equation \eq{friedmann2} in a Taylor series. We get, discarding higher-order terms in $A_{l}$ and $1/\sigma$ (the high energy contributions),

%We observe that in order to recover the braneworld cosmology equations, in the limit where $A_{l} \rightarrow 0$, we must fix the phase constant as $\alpha = \frac{\pi}{2}$. 
%However, in the sense of make a more general analysis for other quantities in which follows, for now we do not fix $\alpha$.  

%It is interesting to note that, in the special case where the energy flux through the horizon is composed by CFT modes, if we consider the validity of the first law of thermodynamics and the temperature \eq{}, the entropy associated with the horizon must be interpreted as 
%entanglement entropy \cite{Bhattacharya:2012mi}.

\begin{equation}
H^{2}  = A(\alpha)\rho^{2} + B(\alpha)\rho + C(\alpha)  \; , \label{friedmann2.2}
\end{equation}

\noindent where

\begin{eqnarray}
A(\alpha) = &&\frac{4\pi}{9}\Big(\frac{3\sin{(\alpha)}}{\sigma} - 2A_{l}\cos{(\alpha)}\Big) , \nonumber \\
B(\alpha) = &&\frac{8\pi}{3}\sin{(\alpha)}, \nonumber \\
C(\alpha) = &&\frac{4\pi}{A_{l}}\cos(\alpha)\; . \label{coefficients}
\end{eqnarray}

%The function $C(\alpha)$ is related with the late time cosmological constant, i.e, the value
%of the cosmological constant when $\rho \sim 0$, given by

%\begin{equation}
%\Lambda_{LT} \sim \frac{3}{2A_{L}}\cos{(\alpha)}. \label{late-time}
%\end{equation}

The Eq. \eq{friedmann2.2} can be written in the form

\begin{eqnarray}
H^{2}  = \frac{8\pi}{3}\rho_{tot}\Big(1 - \frac{\rho_{tot}}{\rho_{c}}\Big) \label{lqc-eq}\;\;,
\end{eqnarray}

\noindent where we have taked $\rho_{tot} = \rho + \Lambda$, with $\Lambda$ as a cosmological constant, and

\begin{equation}
\rho_{c}^{-1} = \frac{1}{6}\Big(2A_{l}\cos{(\alpha)} - \frac{3\sin{(\alpha)}}{\sigma}\Big)\;\;, \label{eq-i}
\end{equation}
\begin{equation}
1 - \frac{2\Lambda}{\rho_{c}} = \sin{(\alpha)}\;\;, \label{eq-ii}
\end{equation}
\begin{equation}
\cos{(\alpha)} = \frac{2A_{l}\Lambda}{3}\Big(1-\frac{\Lambda}{\rho_{c}}\Big) \;\;. \label{eq-iii}
\end{equation}

The Raychaudhuri equation can also be obtained for this case 

\begin{equation}
\dot{H}  = -4\pi(\rho_{tot}+p_{tot})\Big(1 - \frac{2\rho_{tot}}{\rho_{c}}\Big) \;\; , \label{raychaudhuri-eq}
\end{equation}

\noindent where $p_{tot} = p - \Lambda$.

If we take the limit where $A_{l} \rightarrow 0$ in the Eqs. \eq{eq-i}, \eq{eq-ii}, and \eq{eq-iii}, and substitute the results in the Eqs. \eq{lqc-eq} and \eq{raychaudhuri-eq}, we recover the usual braneworld cosmology equations 
\cite{Maartens:2003tw}, as expected (preserving the condition  $\Lambda < \rho_{c}$). 

On the other hand, if we fix $l = \beta l_{AdS}$ in $A_{l}$, for $\beta \neq 0$, we can obtain

\begin{equation}
\rho_{c} = 2\sigma \Big(1+ \frac{\pi}{2\beta^{2}} \Big)^{1/2}\; ; 
\end{equation}

\noindent and

\begin{equation}
\Lambda = 2\sigma\Big[\Big(1+\frac{\pi}{\beta^{4}}\Big)^{1/2} -1  \Big] \; .
\end{equation}

%\begin{equation}
%\alpha = \arccos\Big[\Big(\frac{\pi}{2\beta^{4}+\pi}\Big)^{1/2}\Big]\;.
%\end{equation}

In the limit where we have a small cosmological constant ($\beta \gg 1$) \cite{Weinberg:1987dv, Aghanim:2018eyx}, we also get

\begin{equation}
\rho_{c} \approx 2\sigma > 0\;. \label{r-sigma}
\end{equation}

\noindent From such a result, $\rho_{c}$ assumes the role of a critical energy density in the Eqs. \eq{lqc-eq} and \eq{raychaudhuri-eq}, which have  semiclassical corrections similar to that appears in LQC \cite{Bojowald:2001xe, Ashtekar:2006rx, Taveras:2008ke}.

\vspace{2mm}

\noindent \emph{\textbf{- Polymer structures on a brane and their holographic duals.}}

\vspace{2mm}

In order to investigate the microscopic structure of the bounce, we have that using the techniques introduced by Singh and Soni \cite{Singh:2015jus}, we can obtain from the Raychaudhuri equation \eq{raychaudhuri-eq}, the following Hamiltonian for gravity :

\begin{equation}
\mathcal{H}_{grav} = \frac{-3V}{32\pi  \lambda^2}(2 -  e^{ip\sqrt{\Delta}} -  e^{- ip\sqrt{\Delta}}) \;\;, \label{lqc-hamiltonian}
\end{equation}

\noindent with $p$ as the conjugate momentum  to the volume $V$. 

Moreover, 

\begin{equation}
\lambda = (3/(32\pi \rho_{c}))^{1/2} \;\;; \;\; \Delta  = 6\pi/\rho_{c}\; ,
\end{equation}

\noindent where $\lambda$ and $\Delta$ possess dimensions of length squared, and $\rho_{c}$ is a constant-energy density to be provided by the underlying theory \cite{Singh:2015jus}.

In the present work, $\rho_{c}$ is equal to twice the brane tension, according to equation \eq{r-sigma}. Consequently, we obtain

\begin{equation}
\Delta  = \frac{3\pi}{\sigma} \label{r-delta} \;.
\end{equation}

It is important to observe that, the Hamiltonian \eq{lqc-hamiltonian} is not defined in terms of the conjugate momentum to the volume $p$, but in terms of its complex exponentials. Here, it is a key point since such exponentials consist of holonomies, the building blocks of spin networks in LQG.

%As emphasized in  \cite{Singh:2015jus}, such holonomies are absent in the Hamiltonian structure of usual braneworld cosmology. 

%\footnote{We note that, as has been enphasized in \cite{Singh:2015jus}, in the case of the usual branworld cosmology, the Hamiltonian structure obtained does not allows polymer quantization.}

%In fact, the exponentiation of the observables consist in the first step of the polymer quantization method \cite{Corichi:2007tf}. Consequently, such aproach might be the more natural way to quantize the present theory.

As emphasized in \cite{Singh:2015jus}, the appearance of such holonomies is not possible in the case of the usual braneworld cosmology. But in the present scenario, it makes the microscopic description of the brane spacetime naturally related to the polymer quantization methodology \cite{Corichi:2007tf}. In fact, simply promoting the exponentials in the Hamiltonian \eq{lqc-hamiltonian} to operators, we obtain:

\begin{equation}
\widehat{\mathcal{H}}_{grav}  = \frac{-3V}{32\pi G \lambda^2}[2\mathbb{I} -  \widehat{e^{ ip\sqrt{\Delta}}} -  \widehat{e^{- ip\sqrt{\Delta}}}] \;\;, \label{q-lqc-hamiltonian}
\end{equation}

\noindent where in the equation above, the shift operators $\widehat{e^{\pm ip\sqrt{\Delta}}}$ are defined by their action on the states $\psi_{x_{n}} = e^{ipx_{n}}$, in the momentum representation:

\begin{equation}
\widehat{e^{\pm ip\sqrt{\Delta}}}\psi_{x} = e^{\pm i\sqrt{\Delta} p}e^{i px} = e^{i(x \pm \sqrt{\Delta}) p} = \psi_{x\pm \sqrt{\Delta}}\;\;. \label{v-equation} 
\end{equation}

%\noindent and \textbf{consist in $U(1)$ transformations}. 
%\textbf{(Note that such $U(1)$ transformations are related with the discrete evolution of the geometry on the brane. Is has nothing to do with the $U(1)$ transformations that are performed on the $CFT$  gauge fields on the brane).}

As we can observe, the action of the shift operators will correspond to a finite displacement equals 
to $\sqrt{\Delta}$. Consequently, the Hamiltonian \eq{q-lqc-hamiltonian} gives us that the brane spacetime degrees of freedom can be associated with a  polymer structure, defined by a graph, in the form of a regular lattice

\begin{equation}
\gamma_{\sqrt{\Delta}} = \{x \in \mathbb{R}  \mid x = n\sqrt{\Delta}, \forall n \in \mathbb{Z} \}\;\;.
\end{equation}

\noindent Such polymer structure, which is similar to spin networks in LQC \cite{Mielczarek:2012pf}, brings us a discreteness in the position $x$, with a discreteness parameter $\sqrt{\Delta}$, which from the Eq.\eq{r-delta} is defined by the brane tension.
It imposes superselection rules for the gravitational sector on the brane, in a way that the universe will evolve through discrete increments of the scale factor (or some a
function of it, such as an area or volume).

%\begin{figure}[htb]
%     \centering  % figura centralizada
%      \includegraphics[width=5cm, height=8cm]{fig1.pdf}
%      \caption{(a) In full LQG, the quantum geometry state is described by a spin network, where each edge is labeled by a $SU(2)$ representation (for example, $j$, $k$ and $l$ in the picture). (b) In the present scenario, the quantum states of the brane geometry assume the form of a polymer structure, similar to a spin network in LQC. 
%The (red) loop represents a holonomy, given by the shift operators, around the elementary cell.} \label{fig1}
%\end{figure}

Such superselection rules will also affect the string dynamics in the bulk. It is because, we remember that the brane
couples gravitationally to the bulk through closed
strings, that leaves and cross the brane, and whose couplings, $g_{s}$, are related with the brane tension as $g_{s} \sim 1/\sigma$ \cite{Becker:2007zj}.
In this case, from the Eq. \eq{r-delta}, we obtain

\begin{equation}
g_{s}  \sim \Delta     . \label{delta-g}
\end{equation}

\noindent in a way that, the spectrum of closed string modes in the bulk will be constrained by the discrete spacetime evolution on the brane.

\vspace{2mm}

\noindent \emph{\textbf{- Remarks and conclusions}}

\vspace{2mm}

%The existence of pre-big bang phase(s) in the present scenario may have consequences, for example, for the primordial
%gravitational wave spectrum, for the description of the inflationary period, and for structure formation.

In the present work, we have shown that, in the $AdS/CFT$ context, spacetime degrees of freedom of the gravitational theory on the $AdS$ boundary can be described in terms of polymer structures, similar to LQC spin networks, which correspond to the holographic duals of string states living in the bulk. Such a description of the boundary theory provides us with superselection rules to gravity in a way that the cosmological evolution in a $AdS/CFT$ scenario becomes free from the Big Bang singularity, which is replaced by a bounce.

We note that the quantum gravity effects will make the bounce to occur before the universe gets in a regime below the $AdS$ length scale,
avoiding scenarios where the BHHC should not be valid. In fact, from the Eqs. \eq{r-sigma} and \eq{r-delta}, the discreteness parameter $\sqrt{\Delta}$, that will correspond to the minimal size of the universe in the present scenario, is such that $\sqrt{\Delta} \approx 8,89 \; l_{AdS}$.

Some issues related to the results found out in the present letter deserve more discussion in future works. At first, we must remember that, from the bulk point of view,
the Big Bang singularity is produced by the uncontrolled backreaction of the dilaton field, as it diverges when the closed string coupling vanishes \cite{Bak:2006nh, Engelhardt:2015gla, coment}. However, Eq. \eq{delta-g} tells us that the discrete spacetime evolution on the brane constraints $g_{s}$ to assume only finite nonvanishing values. Consequently, the string modes that would lead to the dilaton divergency and, consequently, to the Big Bang singularity, must be cut out. The future analysis must improve our understanding of how the string dynamics in the bulk is affected by the discrete evolution of the brane spacetime.

Secondly, it is needed to note that the polymer structures found out here differ from those appear in LQC, only because they are defined in terms of the brane tension, and not in terms of the Barbero-Immirzi parameter. Such a fact may be interesting for future discussions of an ancient problem in LQG, the so-called Immirzi ambiguity \cite{Rovelli:1997na}. It is because the brane tension can be dynamically determined \cite{Becker:2007zj}. The present results can match the concern by several authors that a solution to the Immirzi ambiguity problem could stay in a possible dynamical determination of the Barbero-Immirzi parameter \cite{Jacobson:2007uj, Taveras:2008yf, Mercuri:2009zi}.

%In this case, the comparisson of the observational constraints on LQC \cite{Bojowald:2011hd} and braneworld cosmology \cite{} can be interesting.

Thirdly, the semiclassical results presented in the equations \eq{friedmann2}, \eq{lqc-eq} and \eq{raychaudhuri-eq} can be easily extended to the case of non-flat universes. However, since the LQC methods have been shown robust only for the flat universe case \cite{Ashtekar:2007em}, in the present work we can advocate the Big Bang singularity resolution only in this situation. On the other hand, the present results have opened the doors for future connections between LQC and $AdS/CFT$, where more general cases can be analyzed.

At last, based on the present results, a reformulation of the $AdS/CFT$ conjecture can be proposed, where the $4D$ gravitational theory on the brane must be promoted from a semiclassical theory, as prescribed by the BHHC, to a quantum gravity theory. In this case, a possible connection between String theory and LQG can be traced through the following holographic correspondence, relating a gravitational theory in the $AdS_{5}$ bulk to a CFT coupled to gravity on the brane: \\

\vspace{1mm}

\noindent\emph{\noindent   A classical $5$-dimensional spacetime in the $AdS_{5}$ bulk, described by string theory,  must correspond to a quantum 4D braneworld spacetime described by spin networks.}\\

%\textbf{In this way, the $AdS/CFT$ conjecture, as a definition of nonperturbative sting theory, can take LQG as the dual boundary theory.}
%\textbf{In the present context, the thatmal nature of the boundary theory must in rooted in the quantum spacetime structure in the form of spin networks revealed in the present work.}

Further investigations must deepen such connection, by considering for example the relationship between string theory and more general spin network states, as those appear in full LQG. Moreover, the conjecture above can be used to analyze other situations where quantum gravity must be important, as for the case of black holes.

\begin{acknowledgements}
The author acknowledges the anonymous referees for useful comments and suggestions. The author also thanks his family for their support during the preparation of this work.
\end{acknowledgements}


\begin{thebibliography}{30}
\bibitem{coment}
Due the $S$-duality, it is equivalent to the situation where the string coupling goes to infinity.
\end{thebibliography}
\end{document}